\documentclass[conference]{IEEEtran}

\IEEEoverridecommandlockouts
\usepackage{cite}
\usepackage{amsmath,amssymb,amsfonts}
\usepackage{algorithmic}
\usepackage{graphicx}
\usepackage{textcomp}
\usepackage{xcolor}
\def\BibTeX{{\rm B\kern-.05em{\sc i\kern-.025em b}\kern-.08em
    T\kern-.1667em\lower.7ex\hbox{E}\kern-.125emX}}

\usepackage[normalem]{ulem}

\ifCLASSOPTIONcompsoc
    \usepackage[caption=false, font=normalsize, labelfont=sf, textfont=sf]{subfig}
\else
\usepackage[caption=false, font=footnotesize]{subfig}
\fi

\usepackage{amssymb,amsmath,amsfonts,amsthm,bm,mathtools,cuted,bbold}

\usepackage[cmintegrals]{newtxmath}

\usepackage{tabularx}
\usepackage{booktabs}

\usepackage[ruled,vlined,linesnumbered]{algorithm2e}
\SetKwComment{Comment}{/* }{ */}

\newcommand{\mc}[1]{\mathcal{#1}}   
\newcommand{\mb}[1]{\mathbf{#1}}    
\newcommand*\diff{\mathop{}\!\mathrm{d}}    
\DeclareMathOperator*{\argmax}{arg\,max}    
\DeclareMathOperator{\sinc}{sinc}

\newcommand{\DL}{^{\rm \scriptscriptstyle {DL}}} 
\newcommand{\UL}{^{\rm \scriptscriptstyle {UL}}} 

\newtheorem{lemma}{Lemma}

\definecolor{gold}{rgb}{0.85,.66,0}
\definecolor{amaranth}{rgb}{0.9, 0.17, 0.31}

\begin{document}
\bstctlcite{IEEEexample:BSTcontrol} 

\title{
   A Random Access Protocol for RIS-Aided Wireless Communications
}


\author{
    {Victor Croisfelt}, {Fabio Saggese}, {Israel Leyva-Mayorga}, {Rados\l{}aw Kotaba}, {Gabriele Gradoni}, {Petar Popovski}
    \thanks{V. Croisfelt, F. Saggese, I. Leyva-Mayorga, R. Kotaba, and P. Popovski are with the Connectivity Section of the Department of Electronic Systems, Aalborg University, Aalborg, Denmark; e-mail: \texttt{\{vcr,fasa,ilm,rak,petarp\}@es.aau.dk}. G. Gradoni is with the Department of Electrical and Electronics Engineering, University of Nottingham, Nottingham, United Kingdom; e-mail: \texttt{Gabriele.Gradoni@nottingham.ac.uk}.}%
}



\maketitle

\begin{abstract}
Reconfigurable intelligent surfaces (RISs) are arrays of passive elements that can control the reflection of the incident electromagnetic waves. While RISs are particularly useful to avoid blockages, the protocol aspects for their implementation have been largely overlooked. In this paper, we devise a random access protocol for a RIS-assisted wireless communication setting. Rather than tailoring RIS reflections to meet the positions of users' equipment (UEs), our protocol relies on a finite set of RIS configurations designed to cover the area of interest. The protocol is comprised of a downlink training phase followed by an uplink access phase. During these phases, a base station (BS) controls the RIS to sweep through its configurations. The UEs then receive training signals to measure the channel quality with the different RIS configurations and refine their access policies. Numerical results show that our protocol increases the average number of successful access attempts; however, at the expense of increased access delay due to the realization of a training period. Promising results are further observed in scenarios with a high access load.

\end{abstract}

\begin{IEEEkeywords}
Reconfigurable intelligent surface (RIS), random access.
\end{IEEEkeywords}

\IEEEpeerreviewmaketitle

\section{Introduction}
Reconfigurable intelligent surfaces (RISs) {are considered a promising technology to be deployed in the next generation of mobile networks, having the ability to} reflect incident electromagnetic waves in a controllable manner with low power consumption~\cite{Wu2021tutorial,9482474}. Recent works focused predominantly on the physical layer features of wireless communication systems aided by RIS, showing potential benefits in terms of spectral and energy efficiencies~\cite{bjornson2021signalprocessing}. Less attention has been paid to protocol and control signaling aspects, which are crucial for system-level integration and operation of RIS. 

\begin{figure}[t]
    \centering
    \includegraphics[trim={0.5cm 0.8cm 0.0cm 0cm}, clip]{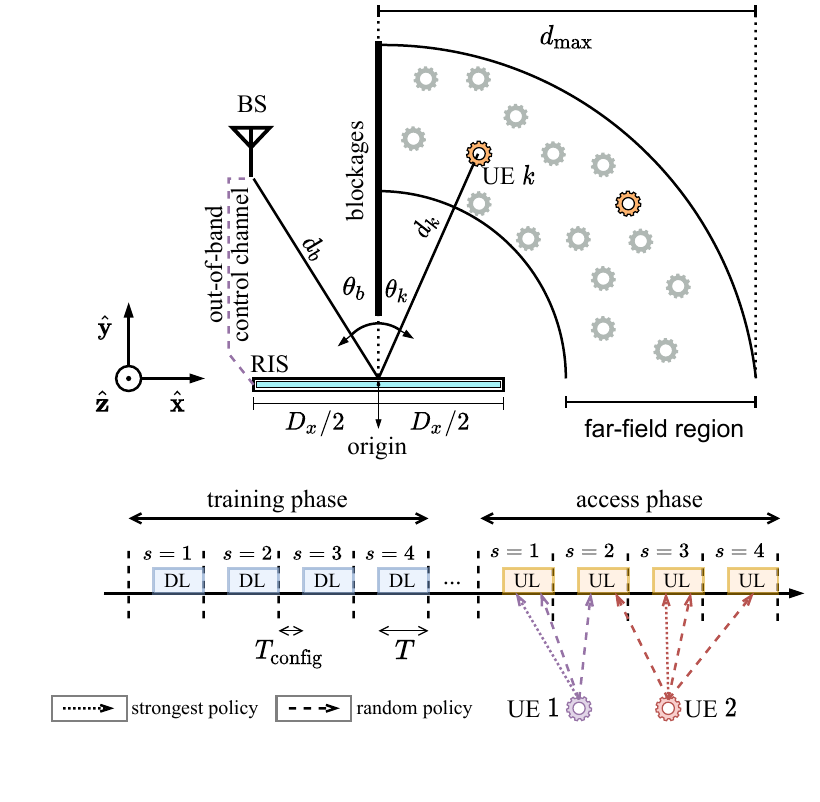}
    \caption{Two-dimensional view of the random access setup assisted by a RIS. {The highlighted UEs are the active ones $k\in\mathcal{K}_a$}. The RIS has negligible thickness, shown for illustrative purposes only. {The proposed protocol is illustrated for $K_a=2$ UEs and $S=4$ configurations, which is comprised of a training phase followed by an access phase. The set $\mathcal{S}$ with cardinality $S$ also enumerates the time slots of duration $T$. The parameter $T_{\text{config}}$ accounts for the time that the RIS cannot be used and, consequently, the BS remains silent. Access policies based on the strongest (single packet) and random (multiple replicas) criteria are illustrated.}}
    \label{fig:system-setup}
\end{figure}

Consider the scenario in Fig.~\ref{fig:system-setup}, where users' equipment (UEs) cannot be served by a base station (BS) due to the blockages. To overcome this problem, a network operator can deploy a RIS to extend the coverage area of the BS and offer network access to the UEs affected by the blockages. In this setting, an important open question is how to design an access protocol for multiple uncoordinated UEs, while taking advantage of the possibility to configure the RIS. There is a gap in the literature regarding the answer of this and related questions. {A recent work}~\cite{Cao2022}, and the references therein, presents designs of medium access control protocol that integrate RISs for multi-user communications. However, these works do not address the problems of initial access and scenarios without an explicit resource allocation, where channel state information (CSI) is unknown. A conventional approach would be to start with a random access that identifies the active UEs, while keeping the RIS configuration fixed, and then tailor the RIS configuration to the scheduled transmissions of the active UEs. This may lead to a significant increase in access delay and, when small data amounts are sent, decrease in overall spectral efficiency. A closely related work is~\cite{Shao2021bayesiantensor}, where the authors considered the activity detection problem for unsourced random access. RIS is used as a means to improve the channel quality and control channel sparsity; however, the work does not clarify how to adequately integrate it into the protocol design.

The contribution of this paper is twofold. 
First, we design a protocol that addresses the main shortcoming of the conventional approach by decreasing the access delay.
The proposed protocol {operates with a finite set of configurations for the RIS}. Each configuration is designed to illuminate a portion of the area of interest where the UEs are located. 
Second, we present a channel model suitable for downlink (DL) and uplink (UL) directions extending state-of-the-art models~\cite{Ozdogan2020pathloss,Ross2021TAP}, which allows to properly calculate the mismatch between the RIS configurations and UEs' positions.


The proposed protocol is divided into a {DL training phase} and an {UL access phase}, as illustrated in Fig.~\ref{fig:system-setup}. 
In the training phase, the RIS (controlled by the BS) sweeps through its configurations, while the UEs estimate the importance, or ``strength'', of each configuration in relation to their positions by relying on training signals sent by the BS. 
The DL training phase is then followed by an UL access phase, where the sweeping is performed again. Active UEs now try to access the network according to access policies that exploit the side-information obtained in the training phase. Thus, the RIS helps to spatially coordinate the access requests from the UEs. 

Numerical results unveil that, despite additional delay to perform the training phase, RIS-aided access coordination pays off notably in crowded scenarios where number of collisions is high. We remark that the proposed protocol is able to provide relevant, prior information to steps following initial access or other network operations. For example, the time slot (RIS configuration) selected by the UE to perform the access attempt is correlated with its position. Therefore, this partial knowledge can be used by the BS to have a rough estimate of the CSI, position, and best configuration for that particular UE, benefiting channel estimation and localization algorithms.

\textit{Notation.} 
Integer sets are denoted by calligraphic letters $\mathcal{A}$ 
with cardinality 
$|\mathcal{A}|$. The circularly-symmetric complex Gaussian distribution is $\mathcal{N}_{\mathbb{C}}(\mu,\sigma^2)$
with mean $\mu$ and variance $\sigma^2$. Lower and upper case boldface letters denote column vectors $\mathbf{x}$ and matrices $\mathbf{A}$, respectively. Identity matrix of size $N$ is $\mathbf{I}_N$ and $\mathbf{0}$ is a vector of zeros. Euclidean norm of $\mathbf{x}$ is $\lVert\mathbf{x}\rVert_2$. Superscript $(\cdot)^*$ denotes complex conjugate. The $\arg\max(\cdot)$ function returns the index of the maximum element of a vector.

\section{System Model}
We consider the random access problem where $K=|\mathcal{K}|$ single-antenna UEs contend for access. The UEs communicate with a single-antenna BS $b$ in a time slotted channel with carrier frequency $f_c$. The system operates in time division duplex mode, where UL and DL communications are carried out in different phases. A phase contains multiple time slots, where each has a duration of $T$ seconds and are comprised of $L$ complex symbols. For simplicity, we assume packet size is fixed and that $L$ symbols are sufficient to transmit it.

Fig.~\ref{fig:system-setup} depicts the considered system setup. The $K$ UEs are uniformly distributed within a predefined area where the line-of-sight (LoS) towards the BS is obstructed by obstacles. The RIS $r$ is placed such that: 1) it has direct LoS towards $b$ and all $k\in\mathcal{K}$ and 2) $b$ and all $k\in\mathcal{K}$ are located in the \emph{far-field region} of the RIS, formally defined in Section~\ref{ss:scattered_signal}. The center of the RIS of size $D_x D_z$ is located at the origin, with $D_x$ and $D_z$ being the RIS dimensions along the $x$- and $z$-axis, respectively. For the sake of simplicity, we consider that BS and UEs are solely located in the $x$-$y$-plane. The angles from the normal line to the RIS towards the BS and towards a UE $k\in\mathcal{K}$ are denoted as $\theta_{b}\in[0,\frac{\pi}{2}]$ and $\theta_{k}\in[0,\frac{\pi}{2}]$, respectively. The distances from the center of the RIS (\emph{i.e.}, the origin) to the BS and to the $k$-th UE are $d_{b}\in\mathbb{R}_+$ and $d_{k}\in\mathbb{R}_+$, respectively. The maximum distance from the origin to the BS and UEs
is denoted as $d_{\max}\in\mathbb{R}_+$. 

The BS controls the operation of the RIS through an out-of-band \emph{control channel} (CC) \cite{bjornson2021signalprocessing}, \emph{e.g.}, a dedicated channel via optical fiber. Consequently, the communication over the CC does not cause interference to the communication between the UEs and the BS.

\subsection{Channel Model} \label{ss:scattered_signal}
The RIS is formed by $N = N_xN_z$ wavelength-scale elements of size $d_x d_z$, that is, $d_x,d_z\leq\lambda$. The elements are indexed by $m\in\{1,2,\dots,N_x\}$ and $n\in\{1,2,\dots,N_z\}$ following the direction of $x$- and $z$-axis, respectively. Each element $(m,n)$ is {assumed to be realized as a metallized layer on a grounded substrate. The metal is assumed to be a perfect electric conductor and able to impress a stable phase shift $\phi(m,n)$ on the incident wave upon moderate amplitude attenuation. As an example, arbitrary phase shifts can be realized with an array of sub-wavelength pixels coupled at sub-wavelength distance, covering a RIS area $\geq\lambda^2$, characterizing a metasurface \cite{Ross2021TAP}. The total size of the RIS} is considered to be much smaller than its distances from the BS and UEs,  that is, the \emph{far-field region} starts from $d_b,d_k\geq\frac{2}{\lambda}\max(D_x^2,D_z^2)$~\cite{balanis2012advanced}. This allows for adopting a uniform plane wave approximation for the RIS scattered field. Considering EM wave polarization, and without loss of generality, we decide to carry on the calculations assuming a transverse magnetic mode that propagates within the plane perpendicular to the $z$-axis (TM$^z$). Having BS and UEs placed in the RIS principal plane leads to the fact that the choice of $\phi(m,n)$ turns out to be independent of the $z$-dimension, since this dimension is orthogonal to wave propagation. Therefore, the phase variation of each element can be simplified as $\phi(m,n) = \phi_s(m)$ for $s\in\mathcal{S}$, where $\mc{S}=\{1,2,\dots,S\}$ enumerates the finite pool of ${S}=\lvert\mc{S}\rvert$ configurations available at the RIS.




In the following, we characterize the \emph{channel coefficients} for DL and UL directions.

\begin{lemma} \label{theo:signals}
Assuming a pure LoS path connecting the BS $b$, the RIS $r$, and the $k$-th UE, the {DL} channel coefficient {is}: 
\begin{equation}\label{eq:receivedsignal}
    \zeta_{k}\DL(s)\in\mathbb{C} = \sqrt{\beta_k\DL} e^{+j \psi_r} \mathrm{A}_{k}(s),%
\end{equation}
where $(s)$ denotes dependency on configurations, and
\begin{equation}\label{eq:pathloss}
    \beta_k\DL\in\mathbb{R}_{+} = \frac{G_{b} G_{k}}{(4\pi)^2} \left(\frac{d_x d_z}{d_{b} d_{k}}\right)^2\cos^2\theta_{b}
\end{equation}
is the DL pathloss with $G_b$ and $G_k$ being the respective antenna gains. Then,
\begin{equation} \label{eq:totalphase}
    \psi_r = -\omega \left(  d_b + d_k - (\sin\theta_b - \sin\theta_k) \frac{N_x+1}{2} d_x \right)
\end{equation}
is the total phase shift (following the path $b\rightarrow{r}\rightarrow{k}$) the signal experiences, with wavenumber $\omega = \frac{2 \pi}{\lambda}$. The array factor arising from the discretization of the RIS into a finite number of elements is: 
\begin{equation} \label{eq:arrayfactor}
    \mathrm{A}_{k}(s) = N_z \sum_{m=1}^{N_x} e^{ j (\omega (\sin\theta_k - \sin\theta_b) m d_x + \phi_s(m))}.
\end{equation}  

\end{lemma}
\begin{proof}
Please refer to appendix~\ref{sec:fields}.
\end{proof}



By convention, {the phases of the DL signals} are considered positive, while UL ones are negative. Hence, for the UL, the channel coefficient $\zeta_{k}\UL(s)$ is computed similarly as in \eqref{eq:pathloss}, but $\beta_k\UL$ considers $\cos^2\theta_{k}$ rather than $\cos^2\theta_{b}$.


%
\subsection{RIS Configurations}
In a random access scenario, the active UEs {and, hence, their exact locations} are not known. Therefore, the RIS configurations are simply designed to cover the area of interest. Each configuration $s\in\mathcal{S}$, is then required to: 1) compensate for the BS position and 2) maximize the received power in a \emph{direction of reflection} in which the RIS is able to ``steer'' the incoming wave within the desired coverage area. By considering the formulation of $\mathrm{A}_{k}(s)$ in~\eqref{eq:arrayfactor}, these two conditions are fulfilled by setting\footnote{This result can also be obtained from the perspective of the Generalized Snell's Law, as motivated in~\cite{Ozdogan2020pathloss}.}
\begin{equation} \label{eq:confoptimal}
    \phi_s(m) = \omega \, m \, d_x (\sin\theta_b - \sin\theta_s),
\end{equation}
where the first part compensates for the BS position while $\theta_s$ represents the direction of reflection. In our case, the coverage area is the first quadrant shown in Fig.~\ref{fig:system-setup} and, hence, $\theta_s\in[0,\frac{\pi}{2}]$. Consequently, we define $\Delta_S=\pi/(2S)$ to be the \emph{angular resolution} of the RIS when $S$ configurations are offered. From there, the set of {desired directions} is defined as\footnote{{This uniform slicing disregards the power profile of RIS reflections. More elaborated configuration designs are left for future work.}}
\begin{equation}
    \Theta(S)=\left\{\theta_s\,\middle|\,\frac{\Delta_S}{2}+(s-1) \, \Delta_S, \forall s\in\mathcal{S}\right\}.
\end{equation}
Then, by substituting~\eqref{eq:confoptimal} into~\eqref{eq:arrayfactor}, we obtain the array factor for the $k$-th UE and a particular RIS configuration {$s$} as
\begin{equation} \label{eq:arrayfactor2}
    \mathrm{A}_{k}(s) = N_z \sum_{m=1}^{N_x} e^{j \omega (\sin\theta_k - \sin\theta_s) m d_x}.
\end{equation}

{
The BS {controls the configuration changes} of the RIS through the CC. We define $T_\text{config}$ to describe how long {the RIS} takes to {switch from a configuration to another}. During this time, the RIS cannot be used, as illustrated by the \emph{silence periods} in Fig.~\ref{fig:system-setup}.
}

\section{RIS-Aided Protocol}\label{sec:random-access-protocol}
In this section, we describe a random access protocol for the RIS-aided communication setup, both illustrated in Fig.~\ref{fig:system-setup}. The description is divided into the definition of a DL training phase and the UL access phase. We assume that each phase is comprised of $S$ time slots, enumerated by the set $\mathcal{S}$, during which a different RIS configuration is active. For simplicity, we assume that each configuration remains active the same amount of time $T$. Moreover, for the protocol to work, the BS needs to send a control packet to the RIS via CC, which contains information about the set of desired directions $\Theta(S)$ and their order. {If the set of directions and/or the order is changed, the RIS needs to sweep the area and inform the UEs of the changes through a new DL training phase. A similar but shorter procedure can be implemented periodically to aid the UEs with synchronization; however, synchronization errors are out of the scope of this paper.} Since we assume an out-of-band CC, RIS control {messages and communication with the UEs can be performed simultaneously without causing interference~\cite{bjornson2021signalprocessing}. However, a silence period of length $T_\text{config}$ is included at the beginning of each slot of both DL and UL phases to wait for the RIS to change} its configuration. Finally, since our focus is primarily on the access phase operation, we consider that the feedback to the UEs occurs flawlessly.




\subsection{Downlink Training Phase}
The DL training phase is mandatory for the UEs wishing to participate in the UL access phase, as it allows them to estimate their \emph{channel qualities} during different RIS configurations (due to channel reciprocity). For that, the BS transmits a training packet at the $s$-th time slot, yielding the following received signal at the $k$-th UE:
\begin{equation}
    \mathbf{w}_k(s)\in\mathbb{C}^{L}=\sqrt{\rho_b}\zeta\DL_k(s)\mathbf{v} + \boldsymbol{\eta}_k,
\end{equation}
where $\rho_b$ is the BS transmit power, $\mathbf{v}\in\mathbb{C}^L$ is a training signal, and $\boldsymbol{\eta}_k\in\mathbb{C}^L\sim\mathcal{N}_\mathbb{C}(\mathbf{0},\sigma^2\mathbf{I}_L)$ is the receiver noise for $k\in\mathcal{K}$ {with variance $\sigma^2$}. After receiving an appropriate number of channel samples for each configuration\footnote{The training phase can be repeated periodically for $\tau\geq1$ times.}, the UEs are able to estimate their DL channel coefficients. Otherwise, the instantaneous channel quality or \emph{strength} of each one of the configurations can be measured by $\lVert\mathbf{w}_k(s)\rVert^2_2$, where the more the configuration aligns with the UE's angle $\theta_s\approx\theta_k$, the larger, or ''stronger'', the absolute value is \emph{expected} to be, as observed from~\eqref{eq:arrayfactor2}. We let $\boldsymbol{\xi}\DL_k\in\mathbb{R}^S=[\lVert\mathbf{w}_k(1)\rVert^2_2,\lVert\mathbf{w}_k(2)\rVert^2_2,\dots,\lVert\mathbf{w}_k(S)\rVert^2_2]^\transp$ denote the collection of channel qualities. Note that, for static UEs and large values of $L$, it is reasonable to assume that UEs perfectly know $\boldsymbol{\xi}\DL_k$ due to the law of large numbers. 

\subsection{Uplink Access Phase}\label{subsection:random-access-protocol}
In the {access phase}, {the BS switches to receive mode, while UEs switch to transmit mode. By assuming that UEs already have a packet to send at the beginning of the access phase, the $K$ active UEs {contend for access considering the} $S$ UL time slots available.} The UEs then need to decide in which slots to attempt access based on theirs \emph{configuration preferences} or \emph{access policies}. Note that, if more than one UE responds in the same slot, a \emph{collision} occurs. Hence, the BS buffers the UL slots to apply a collision resolution strategy, which is discussed in Subsection \ref{subsec:collision-resolution}. In the following, we detail \emph{1)} the access policies and \emph{2)} the collection of slots buffered by the BS. 


%
\subsubsection{Access Policies} 
The decision for a policy takes place before the start of the access phase and is made by each UE in its own. Based on the channel qualities $\boldsymbol{\xi}\DL_k$ obtained during the training phase, we consider two policies in which the RIS assists the $k$-th UE to make its decision on which slots it should use to respond to the BS. We let the set ${\Pi}_k\subseteq\mathcal{S}$ denote the policy adopted by the $k$-th UE. {The policies are illustrated in Fig.~\ref{fig:system-setup} for $K=2$ and $S=4$.}


\noindent \textbf{A) Strongest-configuration policy (SCP):} {Each UE transmits in the time slot $s$ where the associated RIS configuration leads to the best channel quality.} 
That is, $\Pi_k=\argmax_{s\in\mathcal{S}}\boldsymbol{\xi}\DL_k$, where $|\Pi_k|=1,\forall k\in\mathcal{K}$, \emph{i.e.}, each UE sends a \emph{single} packet. 



\noindent \textbf{B) Configuration-aware random policy (CARP):} The $k$-th UE can compute a probability mass function for time slot selection with probabilities proportional to the strength of each configuration used during the respective time slot. Hence, the probability of choosing the $s$-th time slot is:
\begin{equation}
    P(s|{\xi}\DL_k(s))=\dfrac{{{\xi}\DL_k(s)}}{\sum_{s'=1}^{S}{{\xi}\DL_k(s')}}.
\end{equation}
The $k$-th UE then decides if it is going to answer in the $s$-th time slot by tossing a \emph{biased coin} with probability $P(s|{\xi}\DL_k(s))$, consequently defining $\Pi_k$ with $1\leq|\Pi_k|\leq S,\forall k\in\mathcal{K}$. Note that a UE always needs to choose at least one time slot. Different from the strongest policy, each UE can send \emph{multiple replicas} of its packet.


%
\subsubsection{BS buffering}\label{subsubsec:buffered-collisions}
At the $s$-th slot, the BS receives:
\begin{equation}
    \label{eq:rx-invitation}
    \boldsymbol{\psi}(s)\in\mathbb{C}^L= \sum_{k\in\mathcal{K}^{(s)}}^{}\sqrt{\rho_k}{\zeta\UL_k(s)}\boldsymbol{\nu}_k + \boldsymbol{\eta}_b,
\end{equation}
where $\mathcal{K}^{(s)}=\{k|s\in\Pi_k,\forall k\in\mathcal{K}\}$ is the subset of contending UEs which chose to answer in the $s$-th slot, $\rho_k$ is the UE transmit power, $\boldsymbol{\nu}_k\in\mathbb{C}^L$ is the packet content of the $k$-UE with $\lVert\boldsymbol{\nu}_k\rVert^2_2=L$, and $\boldsymbol{\eta}_b\in\mathbb{C}^L\sim\mathcal{N}_{\mathbb{C}}(\mathbf{0},\sigma^2\mathbf{I}_L)$ is the BS receiver noise. In the end, the BS has the collection of received packets $\boldsymbol{\Psi}\in\mathbb{C}^{S\times L}=[\boldsymbol{\psi}(1),\boldsymbol{\psi}(2),\dots,\boldsymbol{\psi}(S)]$ in matrix form.

\subsection{Collision Resolution Strategy}\label{subsec:collision-resolution}
Inspired by contention resolution diversity slotted ALOHA \cite{Popovski2020}, the key idea is to apply SIC-based collision resolution strategies over the buffered packets $\boldsymbol{\Psi}\in\mathbb{C}^{S\times L}$ at the BS. First, we note that the structure of $\boldsymbol{\Psi}$ can be described as an unweighted bipartite graph $B$. This graph is constructed as follows: $B=(\mathcal{K},\mathcal{S},\mathcal{E})$, where $\mathcal{K}$ is the vertex set of contending UEs, $\mathcal{S}$ is the vertex set of {buffered slots}, and $\mathcal{E}$ is the set of edges which can be extracted from $\boldsymbol{\Psi}$; formally, $\mathcal{E}=\{(k,s)|\boldsymbol{\nu}_k \text{ is in } \boldsymbol{\psi}(s), \forall k\in\mathcal{K}, \forall s\in\mathcal{S}\}$. In our scenario, CSI is assumed not available at the BS, 
motivating a simpler implementation of SIC. We describe such method in Algorithm \ref{algo:collision-resolution}, which outputs the \emph{number of successful access attempts}, $\mathrm{SA}$. We let 
$\mathbf{g}_S\in\mathbb{N}^{S}$ 
be the vector
that denote the node degrees of each one of the nodes in the vertex set 
$\mathcal{S}$
. Note that, in practice, the BS does not know $B$ and actually learns how to construct part of it on-the-fly based on the search for \emph{singletons}, that is, non-colliding packets. Moreover, we let $\gamma_{\text{th}}$ be the required signal-to-noise ratio (SNR) for successful decoding.






\begin{algorithm}
    \small
    \caption{Collision resolution strategy}\label{algo:collision-resolution}
    \KwData{$B$, $\boldsymbol{\Psi}$, $\gamma_{\text{th}}$}
    \KwResult{number of successful access attempts, $\mathrm{SA}$}
    initialize $\mathrm{SA}=0$ and compute 
    node degrees 
    $\mathbf{g}_S\in\mathbb{N}^{S}$\;
    \While{\text{a singleton exists, }$1\in\{\mathbf{g}_S\}$ is True}
        {
            find singleton indexes $s^\dagger$ and $k^\dagger$\;
            calculate singleton SNR $\gamma_{\boldsymbol{\psi}(s^\dagger)}$\;
            \If{$\gamma_{\boldsymbol{\psi}(s^\dagger)}\geq\gamma_{\text{th}}$}
            {
            update $\mathrm{SA}\xleftarrow{}\mathrm{SA}+1$\;
            identify other edges with $k^{\dagger}$, $\mathcal{E}^\dagger=\{(k^\dagger,s)|(k^\dagger,s)\in\mathcal{E}\}$\;
            update $\boldsymbol{\psi}(s)\xleftarrow{}\boldsymbol{\psi}(s)-\boldsymbol{\psi}(s^\dagger)$ for other edges\;
            eliminate all edges containing $k^\dagger$, $\mathcal{E}\xleftarrow{}\mathcal{E}\setminus\mathcal{E}^{\dagger}$\;
            }
            update $\mathcal{K}\xleftarrow{}\mathcal{K}\setminus\{k^\dagger\}$, $\mathcal{S}\xleftarrow{}\mathcal{S}\setminus\{s^\dagger\}$, $\mathbf{g}_s$\;
    }
    
\end{algorithm}


\section{Numerical Results}
In this section, we evaluate the effectiveness of the proposed random access protocol for RIS-aided communications by adopting the parameters reported in Table~\ref{tab:simulation-parameters}. For convenience, we assume that $T=1$ and $L$ is large enough to motivate perfect knowledge of 
$\boldsymbol{\xi}\DL_k$ by the UEs. Furthermore, we consider that the protocol is comprised of a single training phase followed by a single access phase. This {is a pessimistic scenario where the training and access phases have the same duration, and, hence, the overhead of training is $50$\% of the total number of slots}. In practice, multiple access phases can be performed after a training phase, depending on the coherence time of the channel. Given that, {the throughput having $S$ configurations is}
\begin{equation}
    \mathrm{th}(S)=\dfrac{\mathrm{SA}(S)}{T_{\text{tr}}(S) + T_{\text{ac}}(S)}\quad\text{[packet/slot]},
    \label{eq:throughput}
\end{equation}
where $\mathrm{SA}$ is the number of successful access attempts obtained from Algorithm \ref{algo:collision-resolution}, $T_{\text{tr}}=T_{\text{ac}}=S(T+T_{\text{config}})$ accounts for the duration of the training and access phases.

As a baseline, {we consider a third access policy in which the UEs {do not measure their channel qualities, that is,} UEs do not perform the training phase. Consequently, their access policies are not coordinated by the RIS. We refer to this baseline as the \textbf{C) unaware random policy (URP)}. This policy is similar to the {CARP} but the slots are chosen based on \emph{fair coin} toss with probability $\frac{1}{S}$. In fact, it can be thought as a framed ALOHA protocol~\cite{Popovski2020} 
but subject to time slots with different channel coefficients. Since the training phase is not considered, the throughput for policy URP does not account for it.}

\begin{table}[htp]
    \centering
    \caption{Simulation parameters}
    \label{tab:simulation-parameters}
    \small
    \resizebox{\columnwidth}{!}{
    \begin{tabular}{cc|cc}
        \hline
        \bf Parameter & \bf Value & \bf Parameter & \bf Value \\ \hline
        carrier frequency, $f_c$ & 3 GHz & UE transmit power, $\rho_{k}$ & 10 mW\\
        max. distance, $d_{\max}$ & 100 m & noise power, $\sigma^2$ & -94 dBm\\
        \# elements along axes, $N_x$, $N_z$ & 10 & BS-RIS distance, $d_b$ & 25 m\\
        element sizes, $d_x$, $d_z$ & $\lambda$ & BS-RIS angle, $\theta_b$ & 45$^\circ$\\
        antenna gains, $G_b$, $G_k$ & $5$ dB & threshold SNR, $\gamma_{\text{th}}$ & 0 dB\\ 
        BS transmit power, $\rho_{b}$ & 100 mW & &\\
        \hline
    \end{tabular}
    }
\end{table}

\begin{figure}[!htbp]
    \vspace{-5mm}
    \centering
    \subfloat[\label{fig:system-num-success} Avg. successful access attempts.]{\includegraphics[width=.48\columnwidth]{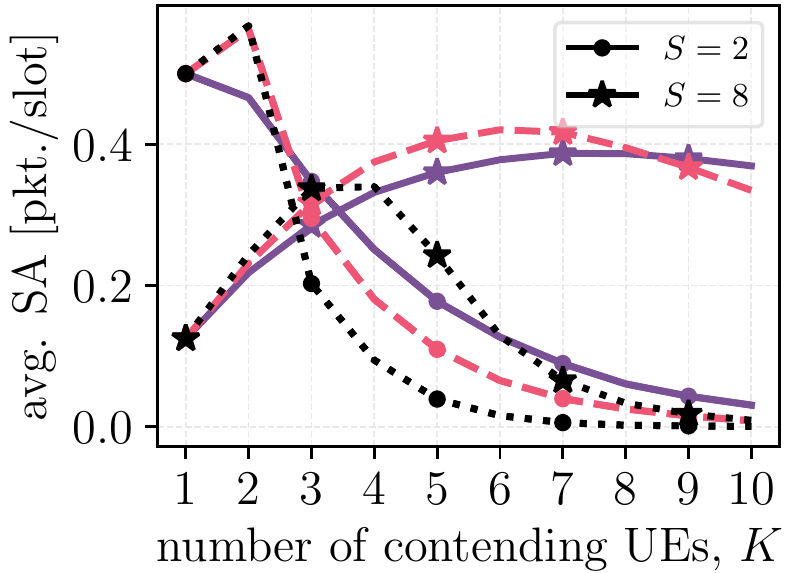}}%
    \hfill
    \subfloat[\label{fig:system-throughput} Optimal avg. throughput w.r.t. $S$. 
    ]{\includegraphics[width=.48\columnwidth]{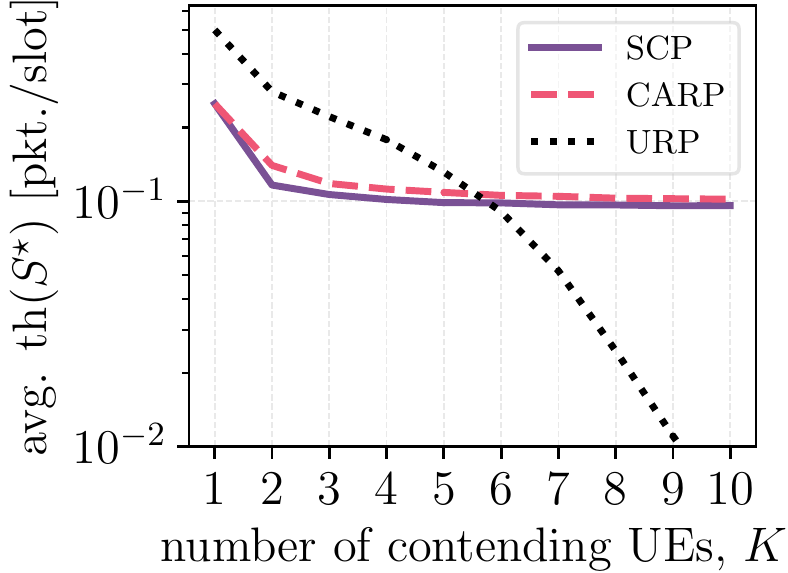}}%
    \caption{Performance of the proposed random access protocol considering $T_{\text{config}}=T$ and different access policies: `$-$' SCP, `$--$' CARP, and `$:$' URP.} 
    \label{fig:performance}
\end{figure}

In Fig.~\ref{fig:performance}, we evaluate the performance of our protocol {as a function of the number of contending UEs trying to access the network, $K$.} 
In particular, Fig,~\ref{fig:system-num-success} shows the average number of successful access attempts $\mathrm{SA}$ for two values of $S$, while Fig.~\ref{fig:system-throughput} presents the optimal throughput w.r.t. the number of configurations $S$, $\mathrm{th}(S^{\star})$.
The results reveal a very important trade-off. On one hand, the use of the RIS to coordinate the access requests from the UEs improves collision resolution, seeing that the average number of success attempts is higher for SCP and CARP. On the other hand, the price to pay for RIS's help is an increased access delay, {introduced by the} 
training phase. Notably, RIS-aided policies outperform when the system becomes crowded (high values of $K$). 
{Between SCP and CARP, which one is better depends on both $K$ and $S$. Higher $S$ can benefit the CARP, exploiting the (possible) multiple access packet replicas sent. 
However, when $K\gg S$, sending a single access packet on the strongest configuration results in better performance (see Fig.~\ref{fig:system-num-success}).}


\section{Conclusion}
We proposed a random access protocol for RIS-aided wireless communication systems. 
The protocol relies on sweeping through a finite set of RIS configurations, 
which improves overall access performance at the price of increased access delay due to the inclusion of a training phase. {Our results show that the overall throughput of the system in high access loads is increased by using the training phase to adapt the access policy of the UEs. However, the overhead of the training phase must be reduced to enhance performance with low access loads.} 


\section*{Acknowledgment}
 This paper was supported in part by the Villum Investigator Grant "WATER" from the Velux Foundations, Denmark and in part by the EU H2020 RISE-6G project under grant number 101017011.

\appendices
\section{Proof of Lemma~\ref{theo:signals}}\label{sec:fields}
Assuming the DL phase and a wave with null phase at the BS, the {generated} electric and magnetic fields are:
\begin{equation}
   \begin{aligned}
    &\mb{E}_b = E_b e^{j \omega (-\sin\theta_b (x-x_b) + \cos\theta_b (y-y_b))} \hat{\mathbf{z}}, \\
    &\mb{H}_b = -\frac{E_b}{\eta} e^{j \omega (-\sin\theta_b (x-x_b) + \cos\theta_b (y-y_b)))} (\hat{\mathbf{x}} \cos\theta_b + \hat{\mathbf{y}} \sin\theta_b),
\end{aligned}
\end{equation}
where $E_b$ is the magnitude of the generated electric field, $\omega$ is the wavenumber, and $\eta$ is the characteristic impedance of the medium.
Using physical optics approximation, the induced current on the $(m,n)$-th element is~\cite{Ross2021TAP}
\begin{equation}
\begin{aligned}
    \mb{J}(m,n)= \frac{2 E_i}{\eta} \cos\theta_b e^{-j \omega \sin\theta_b x} e^{j\phi(m,n)} \hat{\mathbf{z}},
\end{aligned}
\end{equation}
where $E_i = {E_b}/({\sqrt{4\pi} d_b}) e^{j \omega(\sin\theta_b x_b - \cos\theta_b y_b)} $ is the portion of the incident electric field at the RIS independent from the position. By neglecting the border effects, the overall {scattered} field is the superposition of the field scattered by each element, obtaining~\cite[eqs. 6.122a-6.125b p. 290]{balanis2012advanced}
\begin{equation}\label{eq:electricfield}
\begin{aligned} 
    &E_r \simeq E_\phi \simeq 0, \\
    &E_\theta \simeq j\frac{ e^{-j\omega d_k}}{2 \lambda d_k} \eta \sum_{m=1}^{N_x} \sum_{n=1}^{N_z} \int_{A(m,n)} \hspace{-0.6cm}J_z(m,n) e^{j \omega \sin\theta_k x} \diff A,
\end{aligned}
\end{equation}
where $A(m,n)$ denotes the surface of the $(m,n)$-th element. Solving the integral, the scattered field results in
\begin{equation}
\begin{aligned}
    E_\theta = j \frac{e^{j \omega \psi_r}  \mathrm{A}_{k}(s)}{\lambda \sqrt{4\pi} d_k d_r} E_b \cos\theta_b d_x d_z \sinc\left(\frac{d_x}{2} (\sin\theta_k - \sin\theta_b)\right),
\end{aligned}
\end{equation}
where $\mathrm{A}_{k}(s)$ and $\psi_r$ are defined in Lemma~\ref{theo:signals}. 
The relationships between power transmitted and source field, and power received and the effective area at the $k$-th UE are~\cite{Ozdogan2020pathloss}
\begin{equation}
    |E_b|^2 = 2 \eta P_b G_b, \quad A_k^{(e)} = \frac{\lambda^2}{4 \pi} G_k.
\end{equation}
Hence, the overall pathloss is
\begin{equation}
    \frac{P_k}{P_b} = \frac{|E_\theta|^2 A_k^{(e)}}{2 \eta P_b} \simeq \beta_k\DL \mathrm{A}_{k}(s),
\end{equation}
where $\beta_k\DL$ is defined in Lemma~\ref{theo:signals}, and where we approximated $\sinc(d_x(\sin\theta_k - \sin\theta_b)/2) \simeq 1$, since $d_x \ll 1$. Moreover, we note that field amplitudes are normalized to account for power conservation. {The same procedure can be applied for the UL direction, which completes the proof.} \qed



\bibliographystyle{IEEEtran}
\bibliography{main.bib}

\end{document}